\documentclass[mypaper,7pt,twoside]{CoAst}
\usepackage{epsf,graphicx,fancyhdr}
\renewcommand{\chaptermark}[1]{\markboth{#1}{}}

\newcommand{\apj}{ApJ}
\newcommand{\aj}{AJ}  
\newcommand{\mnras}{MNRAS}

\newcommand{\kopf}{\small\itshape Comm. in Asteroseismology\\ Vol. 148, 2006}
\newcommand{\Authors}[1]{\begin{center}\normalsize\bf\sf #1 \end{center}}

\renewcommand{\author}[1]{\begin{center}\normalsize\bf\sf #1 \end{center}}
\newcommand{\Address}[1]{\begin{center}\small\sf #1 \end{center}}

\renewenvironment{abstract}{\section*{Abstract}\normalsize\sf}{}
\newcommand{\References}[1]{\begin{flushleft}{\large References\\}\vspace*{2mm}\small #1 \end{flushleft}}

\newcommand{\chapterDSSN}[2]{\chapter[\sf\normalsize #1\\ \footnotesize \hspace*{5mm}by #2 \sf\normalsize][]{#1\\}\rhead[\fancyplain{}{\sf\footnotesize \center{#1}}]{\fancyplain{}{\sffamily\thepage}}\lhead[\fancyplain{\kopf}{\sffamily\thepage}]{\fancyplain{\kopf}{\sf\footnotesize \center{#2}}}}

\newcommand{\figureDSSN}[5]{\begin{figure}[#4]
\centering
\includegraphics*[#5]{#1}
\caption{#2}
\label{#3}
\end{figure}}

\renewcommand{\chaptername}{}
\newcommand{\acknowledgments}[1]{\vspace*{5mm}\noindent\begin{bf}Acknowledgments. \end{bf} #1}

\begin{document}
\sf
\chapterDSSN{Discovery of hybrid $\gamma$ Dor and $\delta$ Sct
pulsations \\ in BD+18 4914 through MOST spacebased photometry}
{Rowe et al.}
\Authors{J.F. Rowe$^1$, J.M. Matthews$^1$, C. Cameron$^1$, D.A. Bohlender$^7$,\\
H.  King$^1$, R. Kuschnig$^1$, D.B. Guenther$^2$, A.F.J. Moffat$^3$, S.M.
Rucinski$^4$,\\ 
D. Sasselov$^5$, G.A.H. Walker$^1$, W.W. Weiss$^6$}
\Address{$^1$ Department of Physics and Astronomy,
University of British Columbia\\
6224 Agricultural Road, Vancouver BC V6T 1Z1\\
$^2$ Department of Astronomy and Physics, St. Mary's University \\
Halifax, NS B3H 3C3, Canada\\
$^3$ D\'epartement de physique, Universit\'e de Montr\'eal\\
C.P. 6128, Succ.\ Centre-Ville, Montr\'eal, QC H3C 3J7, Canada\\
$^4$ David Dunlap Observatory, University of Toronto\\
P.O.~Box 360, Richmond Hill, ON L4C 4Y6, Canada\\
$^5$ Harvard-Smithsonian Center for Astrophysics \\
60 Garden Street, Cambridge, MA 02138, USA\\
$^6$ Institut f\"ur Astronomie, Universit\"at Wien \\
T\"urkenschanzstrasse 17, A--1180 Wien, Austria \\
$^7$ National Research Council of Canada, 
Herzberg Institute of Astrophysics, \\
5071 West Saanich Road, Victoria, BC V9E 2E7, Canada
}

\noindent
\begin{abstract}
We present a total of 57 days of contiguous, high-cadence photometry
(14 days in 2004 and 43 in 2005) of the star BD+18 4914 obtained with
the MOST\footnote{MOST is a Canadian Space Agency mission, operated
jointly by Dynacon, Inc., and the Universities of Toronto and British
Columbia, with assistance from the University of Vienna.} satellite.
We detect 16 frequencies down to a signal-to-noise of 3.6 (amplitude
$\sim$ 0.5 mmag).  Six of these are less than 3 cycles/day, and the other
ten are between 7 and 16 cycles/day. We intrepret the low frequencies
as g-mode $\gamma$ Doradus-type pulsations and the others as $\delta$
Scuti-type p-modes, making BD+18 4914 one of the few known hybrid
pulsators of its class. If the g-mode pulsations are
high-overtone
non-radial modes with identical low degree $\ell$, we can assign a
unique mode classification of $n$=\{12, 20, 21, 22, 31, 38\} based on
the frequency ratio method.
\end{abstract}

\section{Introduction}
$\gamma$ Doradus stars pulsate with typical periods of about 0.8 days
(Kaye et al. 1999), consistent with high-overtone nonradial g-modes.
They represent one of the newest classes
of pulsating variable stars, and about half of the currently known
$\gamma$ Doradus stars lie within the $\delta$ Scuti instability
strip (Handler 2005). The $\delta$ Scuti variables exhibit p-modes
of low radial order, seen only in low degree photometrically, with
typical periods of a few 
hours.  Handler \& Shobbrook (2002) have shown that the pulsation
characteristics of the two classes can be clearly separated by their
values of the pulsation constant Q.

The overlap in physical properties of $\gamma$ Doradus and $\delta$
Scuti stars suggested the possibility that hybrid pulsators may
exist. The astroseismic implications are exciting since the g-modes
would probe the deep interior of the star and the p-modes, its
envelope. This has lead to photometric monitoring of $\gamma$ Doradus
and $\delta$ Scuti stars to search for hybrid behaviour.  The first
such hybrid to be discovered was in the binary system HD 209295 by
Handler et al. (2002), from careful monitoring of 26 $\gamma$ Doradus
variables (Handler \& Shobbrook 2002) but the $\gamma$ Doradus
pulsations in the primary component are likely caused by tidal
interactions with the secondary.  The first convincing case of a
single hybrid star, the Am star HD 8801, was discovered by Henry \&
Fekel (2005) from monitoring 39 stars from a volume-limited
sample of 114 $\gamma$ Doradus candidates.  HD 8801 shows frequencies
clustered around 3, 8 and 20 cycles/day (c/d).  There were only two
frequencies in the $\gamma$ Doradus range and 4 frequencies
classified as $\delta$ Scuti in nature.  The low number of
frequencies makes this star a challenging subject for
asteroseismic modeling, but just the existence of a hybrid
single star, and the chemical peculiarity of HD 8801, point to
new and interesting astrophysics.

We present photometry of the star BD+18 4914 ($\alpha$ =
$22^h02^m38^s$, $\delta$ = $+18^o54'03''$ [J2000], V=10.6,
B=11.1)\footnote{This research has made use of the SIMBAD database,
operated at CDS, Strasbourg, France.} by the MOST (Microvariability
\& Oscillations of STars) satellite in which we detect frequencies
consistent with hybrid pulsations.

\section{Photometry}

MOST (Walker, Matthews et al. 2003) is a microsatellite housing a
15-cm telescope feeding a CCD photometer through a custom broadband
optical filter.  Launched in June 2003 into an 820-km circular
Sun-synchronous polar orbit (period = 101.413 min), MOST can monitor
stars in its Continuous Viewing Zone (CVZ) for up to 8 weeks without
interruption.  It collects photometry in three ways: (1) Fabry
Imaging, projecting an extended image of the telescope pupil
illuminated by a bright target (see Matthews et al. 2004; Reegen et
al. 2005); (2) Guide Star photometry, based on onboard processing
of faint stars used for telescope pointing (see Walker et al. 2005);
and (3) Direct Imaging, where defocused star images (FWHM $\sim$ 2.5
pixels) are projected onto an open area of the Science CCD. This
last technique was used to obtain the BD+18 4914 photometry, and
details about the Direct Imaging process and reduction are provided
by Rowe et al. (2006b); hereafter, RMSK.

The observations of BD+18 4914 were carried out during 14-30 August
2004 and 1 Aug - 15 Sept 2005, for a total of 57 days, during Direct
Imaging photometry of the transiting exoplanet system HD 209458
(RMSK). Panel D of Figure 1 shows the 2005 observations of BD+18
4914 using 40-min bins to clearly demonstrate the low-frequency
pulsations with periods around 1 day.

\figureDSSN{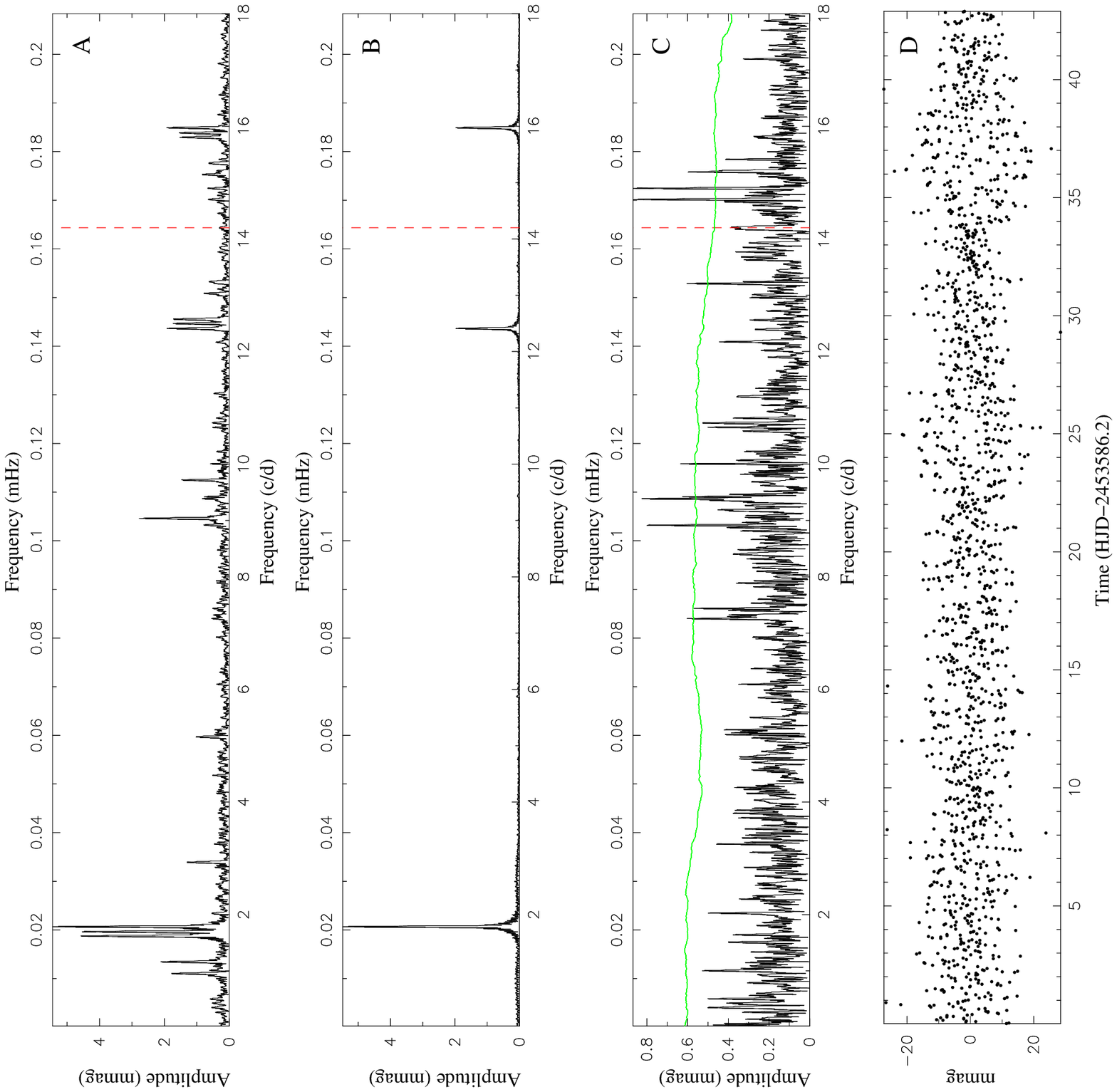}{Panel A shows the amplitude spectrum of the
2005 data set.  Panel B shows the corresponding window function for
the highest peak in Panel A.  Panel C shows the amplutude spectrum
after prewhitening of the 8 strongest frequencies.  The detection
limit of 3.6 times the noise is also shown. (The MOST orbital
frequency is marked as a dashed vertical line.) Panel
D presents the photometric data in 40-min bins to
highlight the low-frequency (2 c/d) oscillations.}
{The FT and Data}{!ht}{clip,angle=270,width=4.5in}

The exposure time was 1.5 sec, sampled once every 10 seconds.  The
2005 data have a raw duty cycle of 97.3\%, with about 360 000
measurments in 43 days.  After rejection of points with extreme cosmic
ray activity and other obvious outliers, the net duty cycle is 75\%.
The 2004 data have a net duty cycle of 70.4\%.  The combined data set
have a total of 443 154
measurements  in 57 days.  There are  gaps about 30 minutes long
during each 101.4 minute orbit of the satellite.  The data do not
suffer from the cycle/day aliases common to groundbased photometry.
They have only minor alias sidelobes at 14.2 c/d (see the spectral
window function in Figure 1) which do not lead to any ambiguities in
the frequency identifications. 

The photometric reduction scheme is described in RMSK and Rowe et
al. (2006, in preparation) for the 
2004 and 2005 data sets, respectively. After dark and flatfield
corrections, photometry is obtained by a combination of aperture
measurments for the core of the stellar point spread function (PSF)
and a fit to the Moffat-profile (Moffat 1969) model for the wings of
the PSF.

The observations are made through a single broadband filter (350 -
750 nm) made specifically for the MOST mission, which has about
$3\times$ the throughput of a Johnson V bandpass but is not tied
to any standard photometric system.

\section{Frequency Analysis}
A preliminary analysis of the 2004 observations of BD+18 4914 was
presented in Rowe et al. (2006a) that showed the dual nature of
its pulsations. Here we take a more methodological approach to the
frequency analysis.

We begin by computing the discrete Fourier transform (DFT) of the
2005 time series. The amplitude spectrum is shown in panel A
of Figure 1. The data are then fitted using an equation of the
form
\begin{equation}
mag = A_0 + \sum _{j=1,n} A_j \cos (2\pi f_j t + \phi_j),
\end{equation}
Where $A_0$ is a linear offset and $f_j$, $A_j$ and $\phi_j$ are
the frequency, amplitude and phase for each successive peak found
in the amplitude spectrum using the Leveberg-Marqardt approach
(Press et al. 1992, p. 678).  After each fit, the DFT of the
residuals is recalculated and the next largest amplitude is
chosen, but only if the Signal-to-Noise (S/N) is greater than 3.6.
The S/N is defined as the amplitude of the peak in the spectrum
divided by the mean of
nearby frequencies.  We use a window about
3 c/d wide in frequency space, centred on the highest peak,
to calculate the mean which we use as an estimate of the local
noise floor.

We detect 16 frequencies in the 2005 photometry with a S/N
greater than 3.6. Our results and best fit parameters are listed
in Table 1.

If the same procedure is repeated for the 2004 photometry, then
the relatively short duration (14 days) of the data set
causes degeneracies in the nonlinear solution because of poor
frequency resolution.  Specifically, the solutions for frequencies
$j=\{2,3\}$ in Table 1 converge to identical values with phases
offset by $\pi$ radians. The amplitudes in turn grow unreasonably
large.  The problem is illustrated in Figure 2. The two top
panels of Figure 2 compare the amplitude spectra of the
2005 and 2004 data sets in the frequency range 0.1 - 3 c/d. 
The 2005 data set has a higher frequency resolution due to its
longer duration.  The 2004 amplitude spectrum shows the
same three peaks as seen in 2005, but the amplitudes of the
peaks appear to be larger in amplitude. This is due to the degeneracy
of the solution 
at these low frequencies in the shorter time series. The
two bottom panels of Figure 2 compare the DFT at a
higher frequency range, 8 - 11 c/d.  The frequency and
amplitude changes seen in the low frequency range are no
longer apparent.

To avoid the degeneracy problem, the frequency solution from the
2005 dataset is
used as the initial solution for the non-linear
routine applied to the 2004 data, with the frequencies held as
fixed parameters to derive the amplitudes to be determined.  If
one examines the DFT of the residuals from the best fit there are
no significant peaks remaining, thus the 2005 frequencies
solution is valid for the 2004 data set. The best fit parameters
are presented in column 5 of Table 1.  This does not imply that
the frequencies are constant from 2004 to 2005, only that the
2004 data set is too short to give meaningful results for a a change
in frequeny in the low frequency regime. 

\figureDSSN{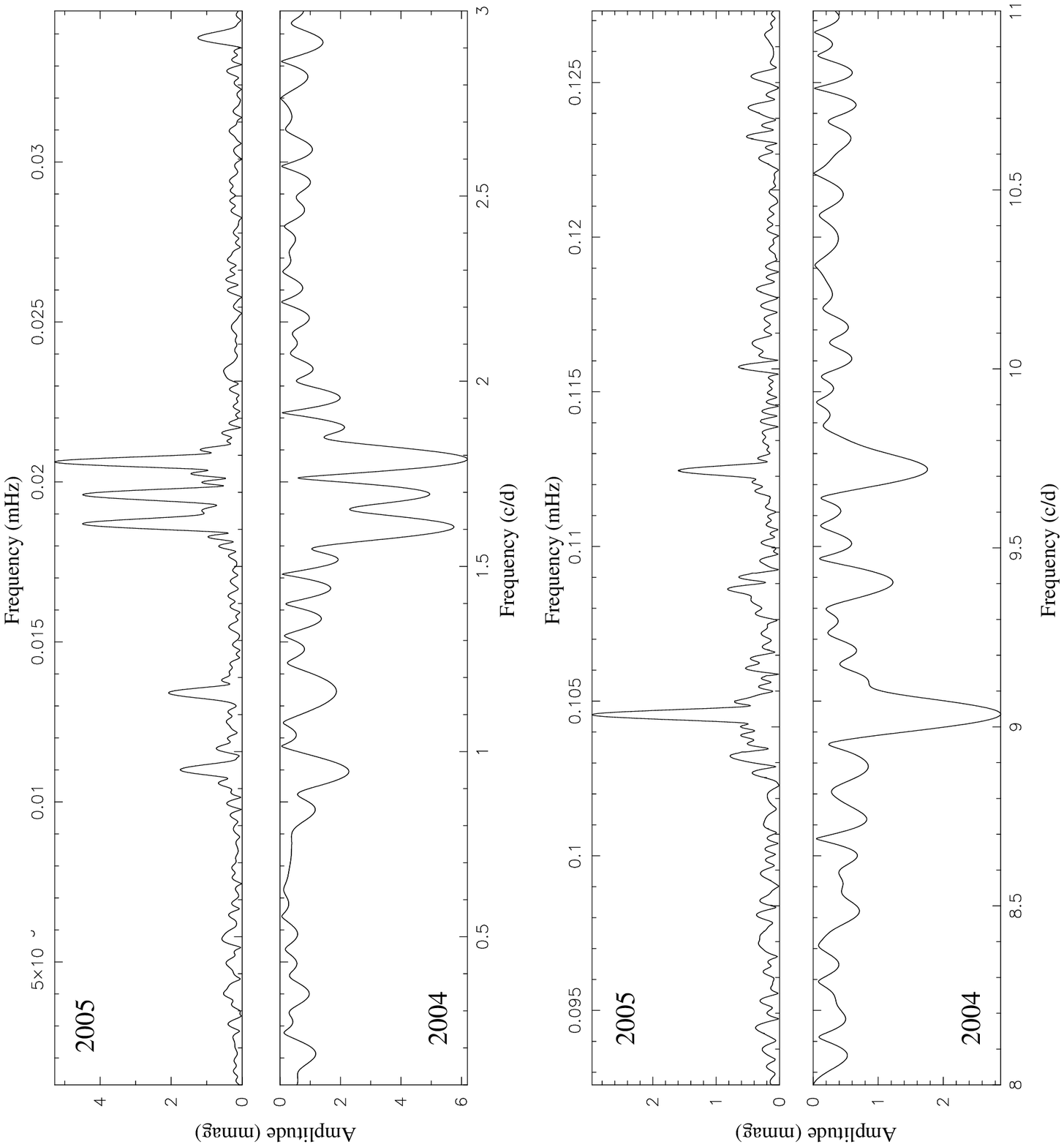}{Comparisons of the 2004 and 2005
Fourier amplitude spectra.  The two top panels
compare the two data sets in the frequency range 0 - 3
c/d. The two bottom panels cover the range 8 - 11 c/d.}
{Comparison of 2004 and 2005 FTs}{!ht}{clip,angle=270,width=4.5in}

To estimate the errors in our fitted parameters, we perform a
"bootstrap" analysis. This method involves redetermining the
fitted parameters with randomly generated data sets.   The
new datasets are created by randomly selecting data from the
original time series with replacement.  In other words,
any individual data point can be chosen more than once but the
total number of selected points is always the same as the
original data set.  The bootstrap method is effective since
it preserves the same noise profile in each random set as
exists in the original data and given enough iterations will
produce error distributions for each fitted variable. For further
demonstrations and discussion of the bootstrap method, we refer
the reader to Cameron et al. (this journal) and Saio et al.
(2006).

We generated 22083 and 18595 bootstrap iterations for the 2004
and 2005 data sets, respectively.  The 1-$\sigma$ error
distributions using a Gaussian model are given in Table 1 for
both data sets.

\begin{table}[ht]
\begin{center}
\begin{tabular}{lcccccc}
\hline \hline $j$ & $f_j$ (c/d)& $A_j$ (mmag)
& $\phi_j$ (rad)
& $A_j^{2004}$ (mmag) & S/N & Q (days)\\
& $\sigma_{f_j}$ & $\sigma _{A_j}$
& $\sigma _{\phi_j}$
& $\sigma _{A_j^{2004}}$ & \\
\hline
 6& 0.9496 & 1.97 & 6.08 & 1.65 & 10.0 & 0.34\\
  & 0.0004 & 0.06 & 0.07 & 0.10 &      & \\
 5& 1.1586 & 2.07 & 3.60 & 1.75 &  9.9 & 0.28 \\
  & 0.0004 & 0.06 & 0.07 & 0.10 &      & \\
 2& 1.6150 & 4.74 & 0.55 & 4.57 & 19.0 & 0.20 \\
  & 0.0002 & 0.06 & 0.03 & 0.10 &      & \\
 3& 1.6924 & 4.60 & 2.05 & 4.48 & 20.0 & 0.19 \\
  & 0.0002 & 0.06 & 0.03 & 0.10 &      & \\
 1& 1.7829 & 5.25 & 5.99 & 4.98 & 19.4 & 0.18 \\
  & 0.0002 & 0.06 & 0.03 & 0.10 &      & \\
 8& 2.9286 & 1.22 & 3.08 & 0.87 &  7.0 & 0.11 \\
  & 0.0007 & 0.06 & 0.12 & 0.10 &      & \\
14& 7.2530 & 0.65 & 0.22 & 1.04 &  3.9 & 0.04 \\
  & 0.0012 & 0.06 & 0.21 & 0.10 &      & \\
15& 7.4354 & 0.53 & 0.03 & 0.42 &  3.8 & 0.04 \\
  & 0.0015 & 0.06 & 0.27 & 0.10 &      & \\
11& 8.9122 & 0.84 & 4.66 & 0.94 &  4.9 & 0.04 \\
  & 0.0010 & 0.06 & 0.17 & 0.10 &      & \\
 4& 9.0348 & 3.09 & 0.43 & 3.08 & 16.3 & 0.04 \\
  & 0.0003 & 0.06 & 0.05 & 0.10 &      & \\
12& 9.3847 & 0.84 & 0.90 & 1.10 &  5.1 & 0.03 \\
  & 0.0010 & 0.06 & 0.17 & 0.10 &      & \\
 7& 9.7156 & 1.56 & 3.95 & 1.85 &  9.4 & 0.03 \\
  & 0.0005 & 0.06 & 0.09 & 0.10 &      & \\
13&10.0043 & 0.60 & 3.09 & 0.50 &  3.9 & 0.03 \\
  & 0.0013 & 0.06 & 0.23 & 0.10 &      & \\
10&14.6977 & 0.91 & 3.07 & 0.64 &  7.0 & 0.02 \\
  & 0.0010 & 0.06 & 0.16 & 0.10 &      & \\
 9&14.8967 & 0.87 & 3.09 & 0.85 &  6.9 & 0.02 \\
  & 0.0010 & 0.06 & 0.16 & 0.10 &      & \\
16&15.4106 & 0.46 & 3.76 & 0.24 &  3.8 & 0.02 \\
  & 0.0018 & 0.06 & 0.31 & 0.09 &      & \\
\hline\\
\end{tabular}
\caption{Observed frequencies and parameters for BD+18 4914.  The
  epoch is HJD=2453586.20349121.} 
\end{center}
\end{table}

\section{A hybrid pulsator}

The frequencies found in BD+18 4914 cluster in the two ranges
typical of $\gamma$ Dor and $\delta$ Sct oscillation modes, making
this a clear candidate for a hybrid pulsator.

We can quantify this assessment using the criterion established
by Handler \& Shobbrook (2002) that the pulsation constant Q
distinguishes the $g-$ and $p-$ modes in this type of star. It was
shown that although the pulsation periods of $\delta$ Scuti and
$\gamma$ Doradus overlap there is a clear separation when Q is
considered (see Figure 9 of Handler \& Shobbrook (2002)).  To
calculate Q, we require basic properties of the star: log $g$,
$M_{bol}$ and $T_{eff}$.  We obtained a 10\AA /mm spectrum with the
1.8m Plasket telescope at the Dominion Astrophysical
Observatory\footnote{Based in part on observations obtained at the
  Dominion Astrophysical Observatory, Herzberg Institute of
  Astrophysics, National Research Council of Canada} 
covering a range of 6473-6716\AA.  Our initial analysis gives values
of $T_{eff}=7250$ K, 
log $g=3.7$ cgs and $M_{bol}=2.5$. Using Equation 1 from Handler \&
Shobbrook (2002), we compute values of Q for each frequency, and
they are presented in Table 1.

With analogy to the Am star hybrid pulsator HD 8801 (Henry \&
Fekel 2005), we assume that all frequencies less than 3.0 c/d are
of $\gamma$ Doradus type and the frequencies higher than 6 c/d
are of $\delta$ Scuti type.
Although we do not have enough information (multibandpass
photometry or spectral line profile variation data) to make
pulsation mode identifications, we can apply the Frequency Ratio
Method (FRM) described by Moya et al. (2005) to the 6 lowest
frequencies.  This assumes that the observed $\gamma$ Dor
pulsations can be described by the asymptotic approximation under
the assumption of adiabaticity and spherical symmetry (Tassoul
1980).  If the modes all share the same degree $\ell$, then the
ratio of the frequencies can be approximated by
\begin{equation}
\frac{\sigma_{\alpha 1}}{\sigma_{\alpha 2}} \approx \frac{n_2 +
1/2}{n_1 + 1/2},
\end{equation}

Under these assumptions, we have searched for 6 overtone $n$
values which satisfy Equation 2, taking an error of $\pm 1.3
\times 10^{-2}$ for calculation of the sets of possible overtones
(Su\'{a}rez et al.
2005).  Restricting our search to overtones up
to and including $n = 60$, we find only one viable solution:
$n = \{12, 20 ,21, 22, 31, $ and $38\}$ for the frequencies
labeled $j = \{8, 1, 3, 2, 5,$ and $6\}$ in Table 1.  If we
search up to $n = 100$, then the density of natural number ratios
compared to our error bounds allows for 78 more solutions.
Regardless, other mode identification methods need to be applied
to restrict the possible values of degree $\ell$.

Observations of other $\gamma$ Doradus pulsators have shown that
the amplitudes can be variable, such as seems to be the case with
9 Aurigae (see Kaye et al. 1997 and references therein).  Using
our results for the best-fit parameters of the 2004 and
2005 photometry, we can examine the possibility of amplitude
changes over a 1-year interval.  In Figure 3 we plot the
measured amplitudes from 2004 versus those from 2005
(see Table 1) with 1$\sigma$ error bars. No significant
amplitude changes have occurred. We define detection of an
amplitude change  as
\begin{equation}
\frac{\Delta A_j}{\sigma} = \frac{A_j - A_j^{2004}}
{\sqrt{\sigma_{A_j}^2 + \sigma_{A_j^{2004}}^2}}
\end{equation}
where the definitions are same as presented in Table 1.  The
distribution of $\Delta A_j / \sigma$ 
appears to be non-Gaussian.  We can test this with an student
T-test and an F-test.  To do so, we generated 10000 sets of 16
random Gaussian deviates and calculated the student T-test
probability and F-test
probability for the distributions to have
similar means to our $\Delta A_j / \sigma$ sample.  Our adapted
criteria (for the samples to differ) is a probability less than
0.0026 (3 $\sigma$).  For the T-test, none of the 10000 sets
were rejected; thus the means are
statistically similar.  For the
F-test, 26.7\% of cases were rejected (i.e., only 27.6\% produced
a probability less than 0.0026). At our chosen 3$\sigma$
threshold, our sample has a Gaussian distribution thus an amplitude change is not
statistically significant.  If we apply the tests to our 2004 and
2005 amplitude distributions and ask if the two distributions
differ, then the student T-test gives a probability of 90.9\% and
the F-test, a probability of 88.3\% that the two have similar
means and variances, respectively.

\figureDSSN{ampchange.ps}{The measured amplitudes for the 2004 and
2005 observation campaigns plotted against each other, with
1$\sigma$ error bars.}
{Comparing 2004 and 2005 amplitudes}{!ht}{clip,angle=270,width=4in}


\section{Conclusions}

We have presented 14 and 43 days of nearly continuous photometry
obtained by the MOST satellite.  From this photometry, we detect
16 frequencies whose amplitudes have $S/N > 3.6$, clustered in
two ranges consistent with $\gamma$ Doradus-type and $\delta$
Scuti-type pulsations.  With 6 frequencies in the $\gamma$ Doradus
range, application of the FRM method assuming a common degree
$\ell$ yields a unique set of radial orders of the pulsations.
Comparison of the 2004 and 2005 data sets show no
statistically significant changes in the pulsation
amplitudes.  However, this star is scheduled to be observed for
a third time by MOST in the fall of 2006 to gain insight into the
stability of the observed frequencies and remove any degeneracies
from our fits.  Groundbased spectroscopy and multicolour
photometry will be necessary to obtain independent mode
identifications to confirm whether the FRM assumptions we have
made are valid and to take advantage of the
potential of BD+18 4914 for asteroseismology.

\acknowledgments
The contributions of JMM, DBG, AFJM, SR, and GAHW are supported by
funding from the Natural Sciences and Engineering Research Council
(NSERC) Canada.  RK is funded by the Canadian Space Agency. WWW
received financial support from the Austrian Science Promotion
Agency (FFG - MOST) and the Austrian Science Funds (FWF - P17580).

\References
{Cameron, C. et al. 2006, this issue.\\
Handler, G. \& Shobbrook, R.R. 2002, \mnras, 333, 251\\
Handler, G. et al. 2002, \mnras, 333, 262\\
Handler, G. 2005, JApA, 26, 241\\
Henry, G.W. \& Fekel, F.C. 2005, \aj, 129, 2026\\
Kaye et al. 1997, DSSN, 11, 32\\
Kaye et al. 1999, PASP, 116, 558\\
Matthews et al. 2004, Nature, 430, 51\\
Moffat, A.F.J. 1969, A\&A, 3, 455\\
Moya et al. 2005 A\&A, 432, 189\\
Press, W.H. et al. 1992, Numerical Recipes in FORTRAN 77 (2nd ed,;
Cambridge: Cambridge Univ. Press)\\
Reegen, P. et al. 2005, MNRAS, 367, 1417\\
Rowe, J.F. et al. 2006a, MmSAI, 77, 282\\
Rowe, J.F. et al. 2006b, \apj, 646, 1241 (RMSK)\\
Rowe, J.F. et al. in preparation\\
Saio, H. et al. 2006, astro-ph/0606712\\
Su\'{a}rez et al. 2005, A\&A, 443, 271\\
Tassoul, M. 1980, ApJs, 43, 469\\
Walker, G.A.H., Matthews, J.M. et al. 2003, PASP, 115, 1023\\
Walker, G.A.H., et al. 2005, ApJ, 635, 77\\}
\end{document}